\documentclass[prd,aps,showpacs,tightenlines]{revtex4}
\usepackage{graphicx}

\newcommand{\bear}{\begin{array}}  \newcommand{\eear}{\end{array}}
\newcommand{\bea}{\begin{eqnarray}}  \newcommand{\eea}{\end{eqnarray}}
\newcommand{\beq}{\begin{equation}}  \newcommand{\eeq}{\end{equation}}
\newcommand{\bef}{\begin{figure}}  \newcommand{\eef}{\end{figure}}
\newcommand{\bec}{\begin{center}}  \newcommand{\eec}{\end{center}}
\newcommand{\non}{\nonumber}

\newcommand{\del}{\partial}  

\newcommand{\bib}{\bibitem}

\def\NPB#1#2#3{Nucl. Phys. {\bf B#1}, #2 (19#3)}
\def\NPBB#1#2#3{Nucl. Phys. {\bf B#1}, #2 (20#3)}

\def\PLB#1#2#3{Phys. Lett. B {\bf #1}, #2 (19#3)}
\def\PLBB#1#2#3{Phys. Lett. B {\bf #1}, #2 (20#3)}

\def\PRD#1#2#3{Phys. Rev. D {\bf #1}, #2 (19#3)}
\def\PRDD#1#2#3{Phys. Rev. D {\bf #1}, #2 (20#3)}

\def\PRLL#1#2#3{Phys. Rev. Lett. {\bf#1}, #2 (20#3)}

\def\PRT#1#2#3{Phys. Rep. {\bf#1}, #2 (19#3)}

\newcommand{\gtrsim}{ \mathop{}_{\textstyle \sim}^{\textstyle >} }
\newcommand{\lesssim}{ \mathop{}_{\textstyle \sim}^{\textstyle <} }
\newcommand{\ds}{\displaystyle}

\begin{document}

\title{Affleck-Dine mechanism with  negative thermal logarithmic potential}
\author{S. Kasuya$^a$, M. Kawasaki$^b$ and Fuminobu Takahashi$^b$}
\affiliation{$^a$ Helsinki Institute of Physics, P.O. Box 64, FIN-00014,
University of Helsinki, Finland\\
$^b$ Research Center for the Early Universe, University of Tokyo,
Tokyo 113-0033, Japan}
\date{\today}

\begin{abstract}
 We investigate whether the Affleck-Dine (AD) mechanism works when the
 contribution of the two-loop thermal correction to the potential is
 negative in the gauge-mediated supersymmetry breaking models. The AD
 field is trapped far away from the origin by the negative thermal
 correction for a long time until the temperature of the universe becomes
 low enough. The most striking feature is that the Hubble parameter
 becomes much smaller than the mass scale of the radial component of the
 AD field, during the trap. Then, the amplitude of the AD field decreases
 so slowly that the baryon number is not fixed even after the onset of
 radial oscillation.  The resultant baryon asymmetry crucially depends on
 whether the Hubble parameter, $H$, is larger than the mass scale of the
 phase component of the AD field, $M_\theta$, at the beginning of
 oscillation.  If $H < M_\theta$ holds,
 the formation of Q balls plays an essential role to determine the baryon
 number, which is found to be washed out due to the nonlinear dynamics of
 Q-ball formation.  On the other hand, if $H > M_\theta$ holds,
 it is found that the dynamics of Q-ball formation does not affect the
 baryon asymmetry, and that it is possible to generate the right amount of
 the baryon asymmetry.
\end{abstract}

\maketitle


\section{Introduction}
\label{sec:introduction} 

Baryogenesis is one of the main issues in the theories of the early
universe. There have been proposed various mechanisms, among which
Affleck-Dine (AD) mechanism~\cite{AD} is a promising candidate in theories
with supersymmetry (SUSY).  One of the features which distinguish
supersymmetric theories from ordinary ones is the existence of flat
directions, of which the AD mechanism makes use.  In fact, there are many
flat directions in the minimal supersymmetric standard model (MSSM), along
which there are no classical potentials. Such a flat direction is
described by a complex scalar field $\Phi$, called AD field. It is crucial
to determine the dynamics of the AD field to estimate the resultant baryon
asymmetry. However, it has been found that the AD baryogenesis is
complicated by the two causes.  One is the thermal
effect~\cite{thermal,thermal2,anisimov}, and the other is the existence of Q
balls~\cite{Kusenko,Enqvist,Kasuya1}.

As for the thermal effects, we must take account of those which appear at
both one-loop and two-loop orders.  The former contribution to the
effective potential of the AD field is the thermal mass term which might
cause the early oscillation of the AD field.  The latter effect, on which
we would like to focus in this paper, induces the logarithmic potential
(which we call thermal log potential hereafter) and hence totally changes
the dynamics if its coefficient is negative. In that case, the AD field
might be trapped far away from the origin for a long time until the
temperature of the universe becomes low enough. In particular, such a trap
is likely to occur in the gauge-mediated supersymmetry breaking (GMSB)
models~\cite{gmsb} where the potential for the AD field becomes flat at
large amplitude. Therefore, we assume the GMSB models in this paper. Due
to the enduring trap by the negative thermal log potential, the dynamics
of the system is so involved that little has been known about whether the
AD baryogenesis really works. Although the sign of the thermal
contribution to the potential at two-loop order depends on the choice of
flat directions, it is expected to be negative for many flat
directions. Hence it is important to consider the case when the negative
thermal log potential is effective.

The other element which we need to incorporate is the existence of Q
balls~\cite{Kusenko,Enqvist,Kasuya1}. After the AD field starts to
oscillate, it experiences spatial instabilities and deforms into
nontopological solitons, Q balls.  Especially, the Q-ball formation is
inevitable for a flat potential as in the GMSB models
~\cite{Kasuya1,Kasuya2,Kasuya3}.  Once Q balls are formed, they absorb
almost all the baryon numbers~\cite{Kasuya1,Kasuya3}. In the usual
picture, the Q-ball formation takes place after the baryon number is
fixed. However, as shown later, the Q-ball formation plays an
essential role to fix the baryon number if the AD field is trapped by
the negative thermal log potential.

In this paper, therefore, we consider the AD baryogenesis with the
negative thermal log potential in the GMSB models, taking into account
the Q-ball formation.  Since the negative thermal log potential
retards the oscillation of the AD field considerably, the dynamics is
quite different from that for the ordinary AD baryogenesis, and we
need to examine the scenario carefully. It is found that the existence
of the Hubble-induced A-term spoils the baryogenesis through the
Q-ball formation, while it is possible to explain the baryon asymmetry
of the present universe in the absence of the Hubble-induced A-term.

\section{Affleck-Dine Mechanism and Q-ball formation}
\label{sec:admech}

In this section we briefly review the Affleck-Dine mechanism and
properties of the Q balls. In the MSSM, there exist flat directions, along
which there are no classical potentials in the supersymmetric limit. Since
flat directions consist of squarks and/or sleptons, they carry baryon
and/or lepton numbers, and can be identified as the AD field. These flat
directions are lifted by both the supersymmetry breaking effects and the
nonrenormalizable operators with some cutoff scale. In the gauge-mediated
SUSY breaking model, the potential of a flat direction is parabolic at the
origin, and almost flat beyond the messenger
scale~\cite{Kusenko,Kasuya3,Gouvea},
\begin{equation}
    V_{gauge} \sim \left\{ 
      \begin{array}{ll}
          m_{\phi}^2|\Phi|^2 & \quad (|\Phi| \ll M_S), \\
          \ds{M_F^4 \left(\log \frac{|\Phi|^2}{M_S^2} \right)^2}
          & \quad (|\Phi| \gg M_S), \\
      \end{array} \right.
\end{equation}
where $m_{\phi}$ is a soft breaking mass $\sim$ O(1 TeV), $M_F$ the SUSY
breaking scale, and $M_{S}$ the messenger mass scale.  Since the gravity
always exists, flat directions are also lifted by the gravity-mediated SUSY
breaking effects \cite{EnqvistMcDonald98},
\begin{equation}
    V_{grav} \simeq m_{3/2}^2 \left[ 1+K
      \log \left(\frac{|\Phi|^2}{M_G^2} \right)\right] |\Phi|^2,
\end{equation}
where $K$ is the numerical coefficient of the one-loop corrections and
$M_G$ is the gravitational scale ($\simeq 2.4 \times 10^{18}$ GeV). This
term can be dominant only at high energy scales because of small gravitino
mass $\lesssim O(1\mbox{ GeV})$.

\begin{figure}
\includegraphics[width=8cm]{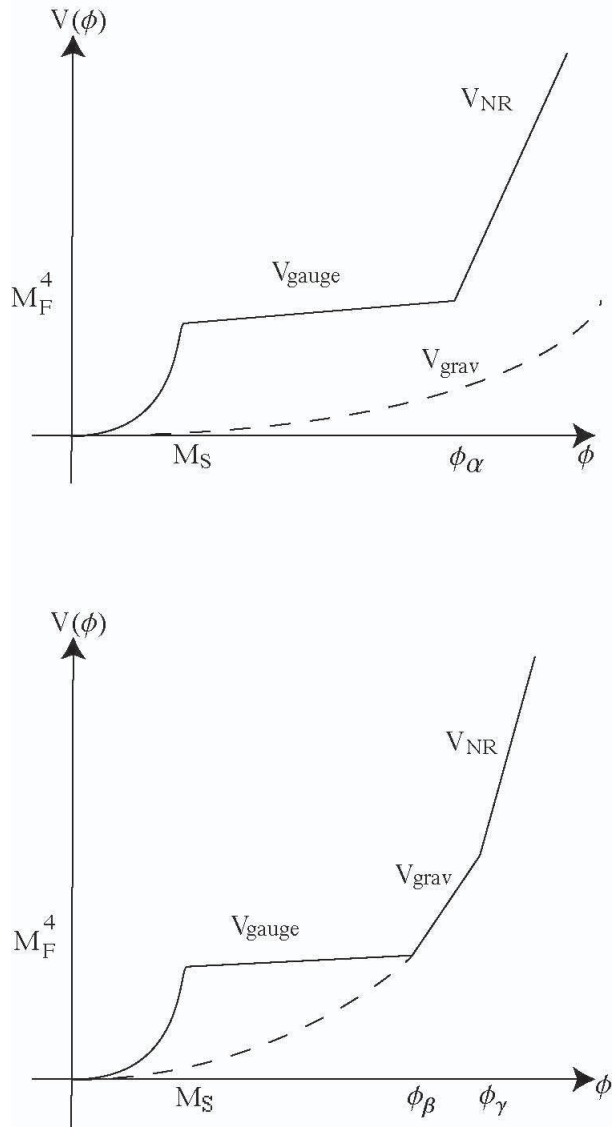} \caption{\label{fig:potential}
Schema for the zero-temperature potential $V_{gauge}+V_{gra}+V_{NR}$.  If
$m_{3/2} < M_F^{\frac{2n-4}{n-1}} M^{-\frac{n-3}{n-1}}$ is satisfied, the
potential takes the form shown in the upper figure.  The
$V_{gauge}$-dominant region connects to the non-renormalizable term
dominant region at $\phi_{\alpha} \simeq \left(M_{F}^{2}
M^{n-3}\right)^{\frac{1}{n-1}}$.  On the other hand, if $m_{3/2} > M_F^{\frac{2n-4}{n-1}}
M^{-\frac{n-3}{n-1}}$ is satisfied, the potential takes the form shown
in the lower figure. The $V_{gauge}$-dominant region connects to the
$V_{grav}$-dominant region at $\phi_{\beta} \simeq
\frac{M_{F}^{2}}{m_{3/2}}$, and then it connects to the non-renormalizable
term dominant region at $\phi_{\gamma}\simeq
(m_{3/2}M^{n-3})^{\frac{1}{n-2}}$.}
\end{figure}

The nonrenormalizable operators, in addition, lift  flat directions.
Assuming that there exist all the nonrenormalizable operators consistent
with both the gauge symmetries of the standard model and the R-parity
in the superpotential, most of flat directions are lifted by the following
operator:
\begin{equation}
\label{eq:spnr}
    W = \frac{1}{n M^{n-3}} \Phi^n\,,
\end{equation}
where $M$ is the cutoff scale. Not only does it lift the flat
potential as
\begin{equation}
\label{eq:pnr}
    V_{NR} = \frac{|\Phi|^{2 n-2}}{M^{2 n -6}}\,,
\end{equation}
but it supplies the baryon number violating A-term as
\begin{equation}
\label{eq:nrA}
    V_{A} = a_m \frac{m_{3/2}}{n M^{n -3}} \Phi^n + {\rm H.c.}\,,
\end{equation}
where $a_m$ is a complex constant of order unity, and we assume the
vanishing cosmological constant. The schema of the zero-temperature
potential $V_{gauge}+V_{grav}+V_{NR}$ is shown in
Fig.~\ref{fig:potential}.

For the AD baryogenesis to work successfully, it is necessary that the AD
field has a large expectation value during the inflationary
epoch~\footnote{Actually, the AD field only has to have a large
expectation value before it starts to oscillate.  However we assumed that
it has a large expectation value during inflation for definite argument.}. To
this end we require the existence of the four-point coupling of the AD
field to the inflaton $I$ in the K\"ahler potential, which leads to the
negative Hubble-induced mass term:
\begin{equation}
\label{eq:hmass}
   V_H = -c_H H^2 |\Phi|^2\,,
\end{equation}
where $c_H$ is a positive constant of order unity.
Also there might be a Hubble-induced A-term, which comes from the
three-point coupling of the AD field to the inflaton in the K\"ahler
potential. We write it as
\begin{equation}
\label{eq:haterm}
   V_{AH} =  a_H \frac{H}{n M^{n -3}} \Phi^n + {\rm H.c.}\,,
\end{equation}
where $a_H$ is a complex constant of order unity.  Though the presence of
this term usually does not change the resultant baryon asymmetry, it does
change the dynamics of the AD field 
when the negative thermal log potential traps
the AD field, as shown later. Since the Hubble-induced A-term is not
necessarily present, we will consider both cases with and without this
term.

Let us consider the thermal effects on the effective potential for the AD
field. During the oscillation of the inflaton, there is  dilute plasma
with  temperature $T \sim (T_{RH}^2 H M_G)^{1/4}$, where $T_{RH}$ is the
reheating temperature. If the AD field directly couples with fields
$\psi_k$ in the thermal bath, it acquires a thermal mass term in the
effective potential at one-loop order:
\begin{eqnarray}
 \label{eq:th1}
     V_T^{(1)} &\simeq& \sum_{f_k |\Phi| < T} c_k f_k^2  T^2 |\Phi|^2,\non\\
               &=&f_{(1)}^2 T^2 |\Phi|^2 {\rm~~~~for~}|\Phi|<f_{(1)}^{-1} T,
\end{eqnarray}
where $c_k$ is a real positive constant of order unity, and $f_k$ is a
Yukawa, or gauge coupling constant between the AD field and
$\psi_k$~\cite{thermal2,afhy}.  For simplicity, we omit the subscript $k$
hereafter with the understanding that $f_{(1)}$ denotes either Yukawa or
gauge coupling constant.  Note that this effect is exponentially
suppressed by the Boltzmann factor when the effective mass of the thermal
particle, $f_{(1)} |\Phi|$, is larger than the temperature.  In addition
to this term, there is another thermal effect on the potential, which
appears at two-loop order, as pointed out in Ref.~\cite{anisimov}. This
comes from the fact that the running of the coupling constant is modified
by integrating out heavy particles which directly couples with the AD
field. This contribution is given by~\cite{afhy2}
\begin{equation}
 \label{eq:th2}
     V_T^{(2)} \sim a_T f_{(2)}^4 T^4 \log\frac{|\Phi|^2}{T^2},
\end{equation}
where $a_T$ is an O(1) constant, and $f_{(2)}^4 = g^4, y^4, y^2 g^2$ ($g$
and $y$ are the gauge and Yukawa coupling constants). The sign of $a_T$
depends on the flat directions, and it is expected to be negative for many
flat directions, for instance, $\overline{u}\overline{d}\overline{d}$
direction, though it is complicated to determine the precise value of
$a_T$. Without specifying flat directions, we would like to focus on the
case with the negative $a_T$ in the following discussion. (The AD
baryogenesis for the positive $a_T$ was considered in
Ref.~\cite{Kasuya3}.)

When the AD field starts coherent rotation in the potential, the baryon
number is generated and fixed immediately in the ordinary AD mechanism,
since the A-terms 
becomes ineffective soon after the oscillation of the AD field. Then the
baryon number density is estimated as
\begin{equation}
 \label{eq:nb}
    n_B(t_{osc}) \simeq \varepsilon \omega \phi_{osc}^2,
\end{equation}
where $\varepsilon(\lesssim 1)$ is the ellipticity parameter, which
represents the strongness of the A-term, and $\omega$ and $\phi_{osc}$ are
the angular velocity and the amplitude of the AD field respectively, at
the beginning of the oscillation (rotation) in its effective potential.
However, this is not the case when the AD field is trapped by the negative
thermal log potential given by Eq.~(\ref{eq:th2}), which will be discussed
in the next section.  In fact, the baryon number is not fixed even after
the AD field starts coherent oscillation, if the Hubble parameter is much
smaller than the mass scale of the radial direction.  It should be
emphasized that Eq.~(\ref{eq:nb}) is valid as long as the baryon number is
fixed soon after the oscillation commenced.

Lastly, we comment on the Q balls. In the course of the coherent
oscillation, the AD field experiences spatial instabilities, and deforms
into nontopological solitons, Q balls~
\cite{Kusenko,Enqvist,Kasuya1}. When
the effective potential is dominated by the zero-temperature term
$V_{gauge}$, the gauge-mediation type Q balls are formed, whose properties
are as follows \cite{Dvali}:
\begin{equation}
    \label{eq:mass}
    M_Q \sim M_F Q^{3/4}, \qquad R_Q \sim M_F^{-1} Q^{1/4},
\end{equation}
where $M_{Q}$ and $R_Q$ are the mass and the size of the Q ball,
respectively.  If the mass per unit charge, $ M_F Q^{-1/4}$, is smaller
than the proton mass $\sim$ 1GeV, the Q ball is stable against the decays
into nucleons, from which it follows that Q balls with very large $Q$ can
be stable.  From numerical calculations~\cite{Kasuya1,Kasuya3}, it is
known that Q balls absorb almost all the baryon charges which the AD field
obtains, and the typical charge is estimated as \cite{Kasuya3}
\begin{equation}
 \label{eq:qcharge}
    Q \simeq \beta \left(\frac{\phi_{osc}}{M_F}\right)^4,
\end{equation}
where $\beta \approx 6 \times 10^{-4}$. Consequently, the present baryon
asymmetry should be explained by the charges which come out of the Q balls
through the evaporation and decay of Q balls.


In the case of the unstable Q balls, they decay into nucleons. The decay
rate is  given by \cite{Coleman}
\begin{equation}
   \left|\frac{dQ}{dt}\right| \lesssim \frac{\omega^{3} A}{192 \pi^{2}},
\end{equation}
where $A$ is a surface area of the Q ball.

In the case of the stable Q balls, the evaporation is the only way to
extract the baryon charges from Q balls. The total evaporated charge
from the Q ball is estimated as \cite{Kasuya3,Laine,Banerjee}, 
\begin{equation}
   \label{eq:chargeevp}
   \Delta Q \sim 10^{15}
   \left(\frac{m_{\phi}}{\mbox{TeV}}\right)^{-2/3}
   \left(\frac{M_F}{10^6\mbox{GeV}}\right)^{-1/3} Q^{1/12}.
\end{equation}
Hence the baryon number density is suppressed by the factor 
$\Delta Q/Q$, in comparison with the case of unstable Q-ball
production.

\section{Dynamics of Affleck-Dine field}
\label{sec:dynm}

In this section, we present a detailed discussion on the dynamics of the
AD field. First we derive the condition that the AD field is trapped by
the negative thermal log potential. In the following argument we set
$|a_m| = c_H = |a_H| = c_k = -a_T = 1$ for simplicity. During inflation
the Hubble parameter is roughly constant, and the initial amplitude of
AD field is set far away from the origin due to the negative
Hubble-induced mass term Eq.~(\ref{eq:hmass}).  The position of the
global minimum is
\begin{equation}
\label{eq:balance}
 \phi \simeq (H M^{n-3})^{1/(n-2)}\,,
\end{equation}
where $\phi \equiv \sqrt{2}|\Phi|$. After inflation ends, the Hubble
parameter decreases, and so does the minimum of the potential. Since the
AD field tracks the instantaneous minimum, Eq.(\ref{eq:balance}) still
holds after inflation. The AD field is trapped by the negative thermal log
potential when $|V_T^{(2)}| > |V_H|$ is satisfied. Here we assumed that
the trap due to the negative thermal log potential occurs while the AD
field tracks the instantaneous minimum given as Eq.(\ref{eq:balance}).
The temperature $T_c$ and the amplitude $\phi_c$ at the beginning of the
trap are
\begin{eqnarray}
\label{eq:trapT}
T_c &=& \left\{
\begin{array}{ll}
\ds{
f_{(2)}^{\frac{n-2}{n}} T_{RH}^{\frac{n-1}{n}} M_G^{\frac{n-1}{2n}} 
M^{-\frac{(n-3)}{2n}}
} & {\rm for~~} \ds{T_{RH} < T_*
,}
\nonumber \\
&\nonumber\\
\ds{
T_*
} & {\rm for~~} \ds{T_{RH}>
T_*,}
\end{array}
\right.\\
\phi_c &=& \left(f_{(2)}^{2}T_c^2 M^{n-3}\right)^{\frac{1}{n-1}},
\end{eqnarray}
where we defined
\beq
T_* \equiv f_{(2)}^{n-2} M_G^{\frac{n-1}{2}} M^{-\frac{n-3}{2}}.
\eeq
It should be noted that $T_{RH} > T_*$ requires a very high reheating
temperature which might be constrained by the gravitino
problem~\cite{Gouvea}. However, if the Q balls dominate the universe, 
they must decay and reheat the universe with the decay temperature,
$T_d \lesssim O(1{\rm GeV})$. Thus the gravitino problem is alleviated due
to the entropy production of the decay of Q balls in this case, and such a
high reheating temperature might be possible.  If it were not for the
gravitino problem, the reheating temperature could be as large as $T_{inf}
\equiv 10^{16}$ GeV. We take this value, because the COBE data implies
that $V_{inf}^{1/4}/\epsilon_{sr}^{1/4} \simeq 6.7\times 10^{16}$ GeV
\cite{LyRi}, where $\epsilon_{sr}< 1$ is the slow-roll parameter, and the
instantaneous reheating is assumed for conservative discussion.

Next we consider when the trap due to the negative thermal log potential
ends.  The AD field starts an oscillation when $|V_T^{(2)}|<V_{gauge}$ is
first satisfied, hence the temperature $T_{osc}$ and the amplitude
$\phi_{osc}$ at that moment are
\begin{eqnarray}
\label{eq:osc}
   T_{osc} &=& f_{(2)}^{-1} M_F\,,\nonumber\\
   \phi_{osc} &=& {\rm min}\left[ \phi_\alpha,~\phi_\beta \right],
\end{eqnarray}
where $\phi_\alpha \equiv \left(M_{F}^{2} M^{n-3}\right)^{\frac{1}{n-1}}$
and $\phi_\beta\equiv M_{F}^{2}/m_{3/2}$. Note that the amplitude of the
AD field, $\phi_{osc}$, at the beginning of oscillation is different from
that in the usual scenario. In the ordinary AD mechanism, the AD field
starts oscillating at $\phi_\alpha$ or $\phi_\gamma \equiv
(m_{3/2}M^{n-3})^{\frac{1}{n-2}}$ depending on the form of the potential
shown in Fig.~\ref{fig:potential}, if one does not
take account of the thermal effects. Such modification is due to the
different dependence of $V_H$ and $V_T^{(2)}$ on $\Phi$.

From the above discussion we can derive the condition that the AD field is
trapped by the negative thermal log potential. It requires the following
inequalities to be satisfied.
\begin{eqnarray}
   \phi_c &>& {\rm Max}\left[\phi_\alpha, \phi_\gamma\right],\\
   f_{(1)} \phi_c &>& T_c,\\
   \phi_{osc} &>& M_S.
\end{eqnarray}
Then we arrive at the following conditions.
\begin{eqnarray}
\label{eq:trap}
  10^{-5} {\rm ~GeV} \lesssim & m_{3/2}& \lesssim 
   m_{3/2}^{(u)},\non\\
 10 {\rm ~TeV} \lesssim &M_F& \lesssim 
   M_F^{(u)}
    \nonumber ,\\ 
  T_{RH}^{(l)} \lesssim& T_{RH}& \lesssim T_{RH}^{(u)}\,,
\end{eqnarray}
where 
$m_{3/2}^{(u)}$, $M_F^{(u)}$, $T_{RH}^{(l)}$ and $T_{RH}^{(u)}$ are given
as
\begin{eqnarray}
   \label{eq:con1}
		m_{3/2}^{(u)} &\equiv& \left\{
								\begin{array}{ll}
						 \ds{1 {\rm ~GeV}} &\ds{
	{\rm for ~~} T_{RH}<T_*},\\
						 &\\
						 	\ds{ {\rm Min}\left[  1 {\rm ~GeV},
		~f_{(2)}^{2n-4}M_G^{n-2}M^{-(n-3)}	 \right]} &
\ds{	{\rm for ~~} T_{RH}>T_*,~ M > f_{(1)}^{\frac{-2}{n-3}} 
                                      f_{(2)}^{\frac{2(n-4)}{n-3}} M_G},
								\end{array}
						\right.\\
 &&\non\\
	M_F^{(u)} &\equiv& 
	\left\{
	 \begin{array}{ll}
	  \ds{ {\rm Min}\left[\sqrt{m_{3/2} M_G},~\left(m_\phi^{n-1} M^{n-3}
	\right)^{\frac{1}{2n-4}}\right] }&\ds{
	{\rm for ~~} T_{RH}<T_*},\\
	  &\\
	  \ds{ {\rm Min}\left[\sqrt{m_{3/2} M_G},~\left(m_\phi^{n-1} M^{n-3}
    \right)^{\frac{1}{2n-4}},~f_{(2)}^{n-1}M_G^{\frac{n-1}{2}}M^{-\frac{(n-3)}{2}}
				   \right] }&
	  \ds{{\rm for ~~} T_{RH}>T_*,~ M >  f_{(1)}^{\frac{-2}{n-3}} 
                                         f_{(2)}^{\frac{2(n-4)}{n-3}}M_G},
	 \end{array}
						\right.\\
 &&\non\\
   T_{RH}^{(l)} &\equiv&
   f_{(2)}^{-2} M_G^{-\frac{1}{2}} \times {\rm Max}\left[ \left(m_{3/2}^{n}
   M^{2(n-3)}\right)^{\frac{1}{2(n-2)}}, ~ \left(M_F^{2n}
   M^{n-3}\right)^{\frac{1}{2(n-1)}} \right] ,\\
 &&\non\\
   T_{RH}^{(u)} &\equiv& \left\{
   \begin{array}{ll}
     T_{inf} & {\rm for~~} \ds{M > f_{(1)}^{\frac{-2}{n-3}} 
                                         f_{(2)}^{\frac{2(n-4)}{n-3}}M_G} \\
     & \\
     {\rm Min}\ds{\left[f_{(1)}^{\frac{n}{n-3}} 
                        f_{(2)}^{\frac{-n+6}{n-3}}
                       M_G^{-\frac{1}{2}} M^{\frac{3}{2}},
     ~~T_{inf}\right]} 
     & {\rm for~~ }\ds{M < f_{(1)}^{\frac{-2}{n-3}} 
                                         f_{(2)}^{\frac{2(n-4)}{n-3}}
                                    M_G},\\
   \end{array}
   \label{eq:con-u1}
   \right. 
\end{eqnarray}
where we do not take account of the gravitino problem, and the bounds for
$M_F$ and $m_{3/2}$ include the constraints due to the scheme of the GMSB.
Notice that $T_{RH}^{(l)}$ comes from $\phi_c > {\rm Max}[\phi_\alpha,
\phi_\gamma]$ and $T_{RH}^{(u)}$ is obtained from the condition $f_{(1)}
\phi_c > T_c$. Also, we must impose the following condition in order to
take the gravitino problem~\cite{Gouvea} into consideration.
\begin{equation}
\label{eq:con-u2} 
 T_{RH} < \gamma \times \left\{
   \begin{array}{ll}
     100{\rm GeV}\sim1{\rm TeV} 
     & {\rm for~~} \ds{2 h^2 {\rm keV} \lesssim
     m_{3/2} \lesssim 100{\rm keV}} \nonumber\\ 
     & \nonumber \\ 
     \ds{10 {\rm TeV}
     \times h^2 \left(\frac{m_{3/2}}{100{\rm keV}}\right)
     \left(\frac{m_{G3}}{1{\rm TeV}}\right)^{-2} } 
     & {\rm for~~ }\ds{m_{3/2} > 100{\rm keV}},
   \end{array}
   \right.
\end{equation}
where $h$ is the present Hubble parameter in units of 100km/sec/Mpc,
$m_{G3}$ is the gaugino mass for $SU(3)_C$, and $\gamma$ reflects the
entropy dilution due to the decay of Q-balls. If Q balls are subdominant
component of the universe throughout, we have $\gamma = 1$. On the other
hand, if Q balls dominate the universe, it is given as
\beq
\label{eq:dilution}
\gamma = \left\{
\bear{ll}
\ds{\frac{T_{RH}}{T_d}} & {\rm for~~}
\ds{T_{eq}>T_{RH}} ,\non \\
&\non\\
\ds{\frac{T_{eq}}{T_{d}}} & {\rm for~~}
\ds{T_{eq}<T_{RH}} ,
\eear
\right.
\eeq
where $T_{eq}$ is the temperature at which the energy of Q balls
becomes comparable to the total energy of the universe, and $T_d$
is the decay temperature of Q balls given as
\beq
 T_d  = \left(\frac{\pi^2 g_*}{90}\right)^{-\frac{1}{4}}
   \sqrt{\frac{M_F M_G}{48 \pi Q^{\frac{5}{4}}}}\,,
\eeq
where $g_*$ counts the effective degrees of freedom of the radiation.

So far we have outlined the dynamics of the amplitude of the AD field.
Now we would like to focus on the motion of its phase, $\theta\equiv {\rm
arg}[\Phi]$, which essentially determines the baryon asymmetry. 
The question we have to ask here is when the baryon number is fixed.  For
this purpose let us begin our analysis by following the evolution of the
baryon number density.  For a moment, we do not consider the effect of the
Q-ball formation.  The equations of motion for $\phi$ and $\theta$ are
\begin{eqnarray}
   \ddot{\phi} + 3 H \dot{\phi} + \frac{\del V}{\del \phi}
   -\dot{\theta} \phi^2 &=&0,\\
   \ddot{\theta} + \left(3 H + 2\frac{\dot{\phi}}{\phi}\right)\dot{\theta}
   +\phi^{-2} \frac{\del V}{\del \theta}&=& 0\,.
\end{eqnarray}
The baryon number density $n_B$ is given as
\begin{eqnarray}
   n_B &=& i q \left(\dot{\Phi}^{*} \Phi- \Phi^{*}\dot{\Phi}\right),
   \nonumber\\
    &=& q \dot{\theta} \phi^2\,,
\end{eqnarray}
where we defined that the baryon charge of the AD field is $q$.
Its evolution is governed by the following equation,
\begin{equation}
\label{eq:ev_of_nb}
    \dot{n_B} + 3 H n_B + q \frac{\del V}{\del \theta}=0.
\end{equation}
The first and second terms mean that the number density is diluted by the
expansion of the universe, and it is the third term that represents the
generation of the baryon number.  Note that the only A-term, $V_A
(+V_{AH})$, can contribute to the third term.  

In the first place we review how the baryon number is fixed in the
ordinary AD mechanism.  In this case, the AD field starts the oscillation
when the Hubble parameter becomes equal to the mass scale of the radial
direction, $M_\phi$. Thereafter the amplitude $\phi$ is reduced by the
Hubble frictional term. Accordingly the magnitude of the A-term decreases
rapidly, since it is a positive power of $\phi$.  Then it can be shown
that the A-term cannot generate the baryon number efficiently after the AD
field starts the oscillation as follows. While the AD field tracks the
slowly decreasing minimum given as Eq.~(\ref{eq:balance}), the motion of
$\theta$ is strongly damped by the frictional term because
$\dot{\phi}/\phi \sim H$ and $H\sim M_\phi \gtrsim M_\theta$, where
$M_\theta$ is the mass scale of $\theta$. The second equality is
satisfied if the Hubble-induced A-term exists. Hence the characteristic
time scale for $\theta$ is $(M_\theta^2/H)^{-1}$, and the baryon number
density is expressed as
\begin{equation}
\label{eq:nbrel}
  n_B = q \dot{\theta} \phi^2 \sim q \frac{M_\theta^{2}}{H} \phi^2\,.
\end{equation}
Then the evolution equation for $n_B$ is given as
\begin{eqnarray}
\label{eq:nbgene}
  \dot{n_B}+3 H n_B &=& -q \frac{\del V}{\del \theta},\nonumber\\
  &\sim& q M_\theta^2 \phi^2 \nonumber,\\
  &\sim& H n_B\,,
\end{eqnarray}
where we used Eq.~(\ref{eq:nbrel}) in the last equality.  From
Eq.(\ref{eq:nbgene}), we can see that the baryon number continues to be
critically generated while the AD field is trapped by the negative Hubble
mass term. That is to say, the decrease of the baryon number density due
to the expansion of the universe is compensated by the baryon number
violating term to some extent, as long as the AD field tracks the
instantaneous minimum given as Eq.~(\ref{eq:balance}).  However, once it
starts the oscillation, the magnitude of the A-term is significantly
reduced, and also the time scale for $\theta$ becomes much smaller
: $\dot{\theta}^{-1} \sim M_\phi^{-1} \lesssim H^{-1}$.  Thus the generation of the baryon
number stops soon after the AD field starts oscillating in the ordinary AD
mechanism.

\begin{figure}
\includegraphics[width=8cm]{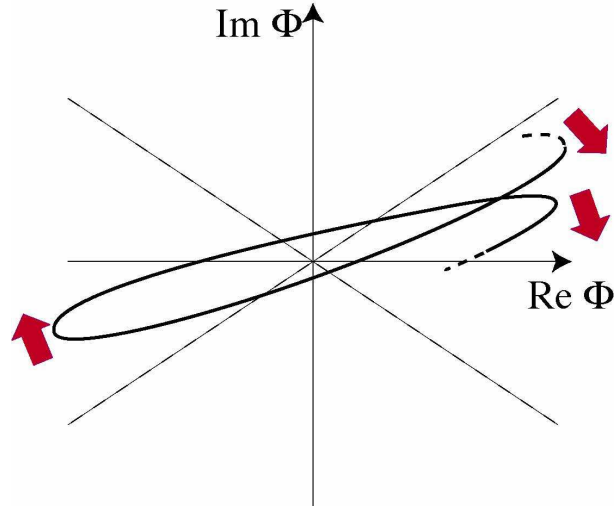} \caption{\label{fig:adosc} This
figure illustrates how the AD field behaves after the beginning of the
oscillation, when it is trapped by the negative thermal log potential. The
thick solid line is the orbit of the AD field.  The dashed lines represent
the hills of the A-term, and the AD field oscillates in the valley
surrounded by them. At each oscillation, it is kicked into the direction
shown as the broad arrow.}
\end{figure}

The usual picture outlined above does not apply to the case when the AD
field is trapped by the negative thermal log potential. First we assume
the existence of the Hubble-induced A-term, and see how it spoils the
baryogenesis.  The relations among the mass scales, $M_\phi$, $M_\theta$
and $H$, are then $M_\phi\sim M_\theta \sim H$ before the trapping due to
the thermal effect.  Once the AD field is trapped by the negative thermal
log potential, its amplitude, $\phi \sim (f_{(2)}^2 T^2
M^{n-3})^{1/(n-1)}$, decreases relatively slowly. While both $M_\phi$ and
$M_\theta$ are decreasing functions of slowly varying $\phi$, the Hubble
parameter diminishes faster.  Thus, $M_\phi \gtrsim M_\theta \gg H$ holds
until the trap ends. Since the mass scale of $\theta$ is larger than the
Hubble parameter,
$\theta$ starts oscillating around its minimum which is determined by the
superposition of $V_A$ and $V_{AH}$, as soon as the AD field is
trapped. During the trap, the amplitude of the oscillation of $\theta$
decreases continuously \footnote{Though the Hubble-induced A-term,
$V_{AH}$, might become smaller than the A-term, $V_A$, it does not
increase the amplitude of $\theta$.}, so that $\theta$ stays very close to
the bottom of the valley of the A-term when $\phi$ starts an
oscillation. Therefore the AD field rotates the orbit with very large
oblateness, as shown schematically in Fig.~\ref{fig:adosc}. It should be
noted that the amplitude of the oscillation of $\phi$ remains almost
constant due to the small Hubble parameter, which has a decisive influence
on the evolution of the baryon number density. In order to follow the
evolution of $n_B$, we estimate the third term in Eq.~(\ref{eq:ev_of_nb}),
which is now given as
\begin{eqnarray}
   q \frac{\del V}{\del \theta} &\sim&q M_\theta^2 \phi^2 
   = \dot{\theta}^{-1} M_\theta^2 n_B,
   \nonumber\\
   &\lesssim& M_\theta n_B\,, 
\end{eqnarray}
where the last equality holds at the end point of each
oscillation. Therefore the baryon number density continues to be generated
around the end point, since $M_\theta \gg H$. Put simply, the valleys of
the A-term become effectively deeper due to the enduring trap, which
causes the intermittent generation of the baryon number.

Consider now the implication of the Q-ball formation. The AD field
experiences instabilities when it oscillates in the $V_{gauge}$ dominated
potential, and deforms into the gauge-mediation type Q balls.  Since the
growth rate of the instability is of the order of $ M_\phi \sim
M_F^2/\phi_{osc}$~\cite{Kasuya3}, Q balls are formed soon after the AD
field starts a radial oscillation~\footnote{
The existence of the A-term does not prevent the formation of Q balls,
because it does not alter the flatness of the potential at all, namely
${\rm Max}[V_{A},V_{AH}] < V_{gauge}$.  In other words, if only a small
part of the energy for the radial component is converted to that for the
phase direction, the AD field can easily get over the hill of the
A-term. Also we confirmed it with the numerical calculation.}.
The most important part of this argument is that the baryon number is not
fixed during the process toward the Q-ball formation. When the fluctuation
becomes comparable to the homogeneous mode, the system goes into the
nonlinear stage, and the AD field is kicked back and forth at random in
the phase direction by the A-term. It is not until Q balls are formed that
the baryon number is fixed.  The reason is as follows. It is known that Q
balls absorb almost all the baryon number. Inside Q balls, the AD field
rotates a circular orbit, so the average of the third term in
Eq.~(\ref{eq:ev_of_nb}) over a round of the orbit vanishes:
\begin{equation}
   \oint_0^{2 \pi} d\theta \frac{\del V}{\del \theta} = 0,
\end{equation}
which means the baryon number is fixed.  Since the baryon number evolves
indiscriminately until the Q-ball formation is completed, the resultant
baryon asymmetry is affected by the highly nonlinear physics of the Q-ball
formation. In rough estimate, the created baryon number before the system
goes into the nonlinear stage is
\begin{equation}
\label{eq:ini_nb}
n_{B}^{(i)} \simeq \frac{c_1}{M_\phi} M_\theta^2 \phi_{osc}^2 
\sin{n \theta_{osc}},
\end{equation}
where $\theta_{osc} (\ll 1)$ is the phase amplitude at the start of the
radial oscillation, and $c_1 M_\phi^{-1}$ with $c_1 \sim
\log{(\phi/\delta \phi)} = O(10)$ is the time for the fluctuation to
become nonlinear. When the fluctuation of the phase becomes large, {\it
i.e.}, $\delta \theta/\theta \sim O(1)$, the AD field moves beyond the
potential hill in the phase direction and the baryon number production
restarts with random phase, $\theta_i({\bf x}) \sim O(1)$. Here $\theta_i$
depends on the position ${\bf x}$.  
Hence this process induces the fluctuation
of the baryon number density, and it takes place over and 
over again until Q-ball formation is completed.  If the process lasted eternally,
however large the baryon asymmetry is, it  would be  completely erased due to the baryon
number violating process in equilibrium.  Note that the baryon number is odd under
the CPT transformation.  In fact, this process is effective for a finite time,
so that  the only finite baryon asymmetry is annihilated.
We can estimate the magnitude of the baryon asymmetry, which is erased
due to the Q-ball formation processes, by calculating 
the typical value of the fluctuation of the baryon number density :
\begin{eqnarray}
\label{eq:nb_fluc}
\delta n_B({\bf x})  &\simeq& \frac{1}{M_\phi} M_\theta^2 \phi_{osc}^2 
\sum_i^{c_2}\sin{n \theta_{i}({\bf x})} , \non\\
&\sim& \pm \frac{\sqrt{c_2} 
}{M_\phi} M_\theta^2 \phi_{osc}^2 ,
\end{eqnarray}
where $c_2 \sim O(10)$ is the number of the restarts, and 
 the contribution of the random phase $\theta_i$ is
evaluated as random walk with $c_2$ steps in the second line. 
Since $\theta_{osc} \ll 1$ in the present situation, we have $n_B^{(i)} \ll \delta n_B$, which
means that 
$n_B$ is dominated by 
the fluctuation due to the Q-ball formation process. Moreover, the dynamics is
chaotic in the sense that a tiny change results in a quite different baryon number.
Thus,  the baryon asymmetry generated during the homogeneous
evolution  is swallowed up in the large fluctuation, leading to the
baryon symmetric universe.

\begin{figure}
\includegraphics[width=8cm]{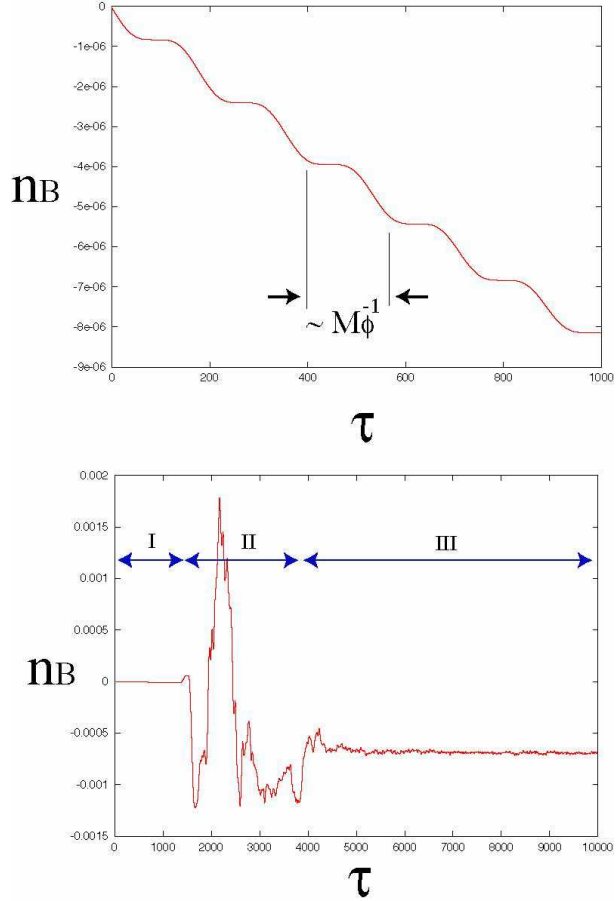} \caption{\label{fig:nbev} The
evolution of the baryon number during Q-ball formation is shown. As shown
in the lower figure, it can be classified into three parts. First the AD
field is kicked around the endpoint at each oscillation (see
Fig.~\ref{fig:adosc}), leading to the intermittent baryogenesis (interval I).
Then the behavior of the baryon number becomes chaotic, as the
fluctuations grows (interval II). Finally, it is fixed after the Q-ball
formation is completed. The upper figure is the blowup of the evolution
for the baryon number during the interval I.
}
\end{figure}

We performed the numerical simulation in order to confirm whether the
baryon asymmetry is really determined through the Q-ball formation. For the
numerical calculation, we take the variables dimensionless as follows.
\begin{eqnarray}
   \varphi &=& \frac{\Phi}{M_S},\non\\
   \tau &=& \frac{M_F^2}{M_S} t,\non\\
   \chi_i &=&  \frac{M_F^2}{M_S} x_i,
\end{eqnarray}
The initial conditions of the homogeneous parts are taken as
\begin{eqnarray}
   {\rm Re}\left[\varphi(0)\right] & =& 1.0\times 10^2,\non\\
   {\rm Im}\left[\varphi(0)\right] & =& 1.0\times 10^{-3},\non\\
   \varphi'(0) &=& 0.0,
\end{eqnarray} 
where the prime denotes the derivative with respect to $\tau$, and we
redefined the phase of the AD field so that the real direction coincides
with one of the valleys of the A-term with the mass scale $M_\theta/M_\phi
= 10^{-1.8}$.  Taking account of fluctuations which originate from the
quantum fluctuations, we give $\varphi$ fluctuations with amplitudes
$10^{-7}$ times smaller than the homogeneous mode. We have confirmed that
the smaller fluctuations just delay the formation of Q balls. For
simplicity we neglected the cosmic expansion. The results are shown in
Fig.~\ref{fig:nbev}. It shows that the baryon number is generated at the
end point of each oscillation for some time $\sim c_1 M_\phi^{-1}$, but it
soon goes into the chaotic phase until the Q-ball formation is
completed. If we change the initial seed for the fluctuation, it is
probable that the resultant baryon asymmetry inside the box of the
lattices becomes positive. The reason why the resultant baryon number
takes nonzero value is that the size of the box under consideration is
finite. If the same calculation were done with the infinitely large box,
the resultant baryon number would be exactly zero.

Lastly we comment on the case without the Hubble-induced A-term. 
Now $M_\phi > H > M_\theta$ at the onset of the oscillation can be
realized since the mass scale of $\theta$ is smaller than that in the
previous case.
Then the initial phase amplitude $\theta_{osc}$ is comparable to the random
phase $\theta_i$, hence the net baryon number can be nonzero. In the next
section we estimate the resultant baryon asymmetry in this case, taking
account of the Q-ball formation.

\section{Baryogenesis}
\label{sec:baryo}

As we have shown in the previous section, the Hubble-induced A-term spoils
the baryogenesis through the Q-ball formation. On the other hand, if it
were not for the Hubble-induced A-term, we can avoid this problem since
$M_\phi > H > M_\theta$ at the beginning of the oscillation is possible.
Then the baryon number just before the Q-ball formation is completed is
roughly estimated as
\begin{equation}
\label{eq:ini_nb2}
n_{B}^{(f)} \sim 
 \frac{c_1}{M_\phi} M_\theta^2 \phi_{osc}^2 
> \frac{ \sqrt{c_2} }{M_\phi} M_\theta^2 \phi_{osc}^2 ,
\end{equation}
where $c_1 \sim c_2 \sim O(10)$,
and we used $\theta_{osc} \sim O(1)$ in
the present case.
Thus we expect that the finite baryon asymmetry remains even if the AD field gets 
randomly kicked in the
phase direction during the nonlinear stage.  In other words, the resultant
baryon number is almost determined due to the dynamics before the AD field
goes into the nonlinear phase, so subsequent Q-ball formation does not
affect the baryon asymmetry.

\begin{figure}
\includegraphics[width=15cm]{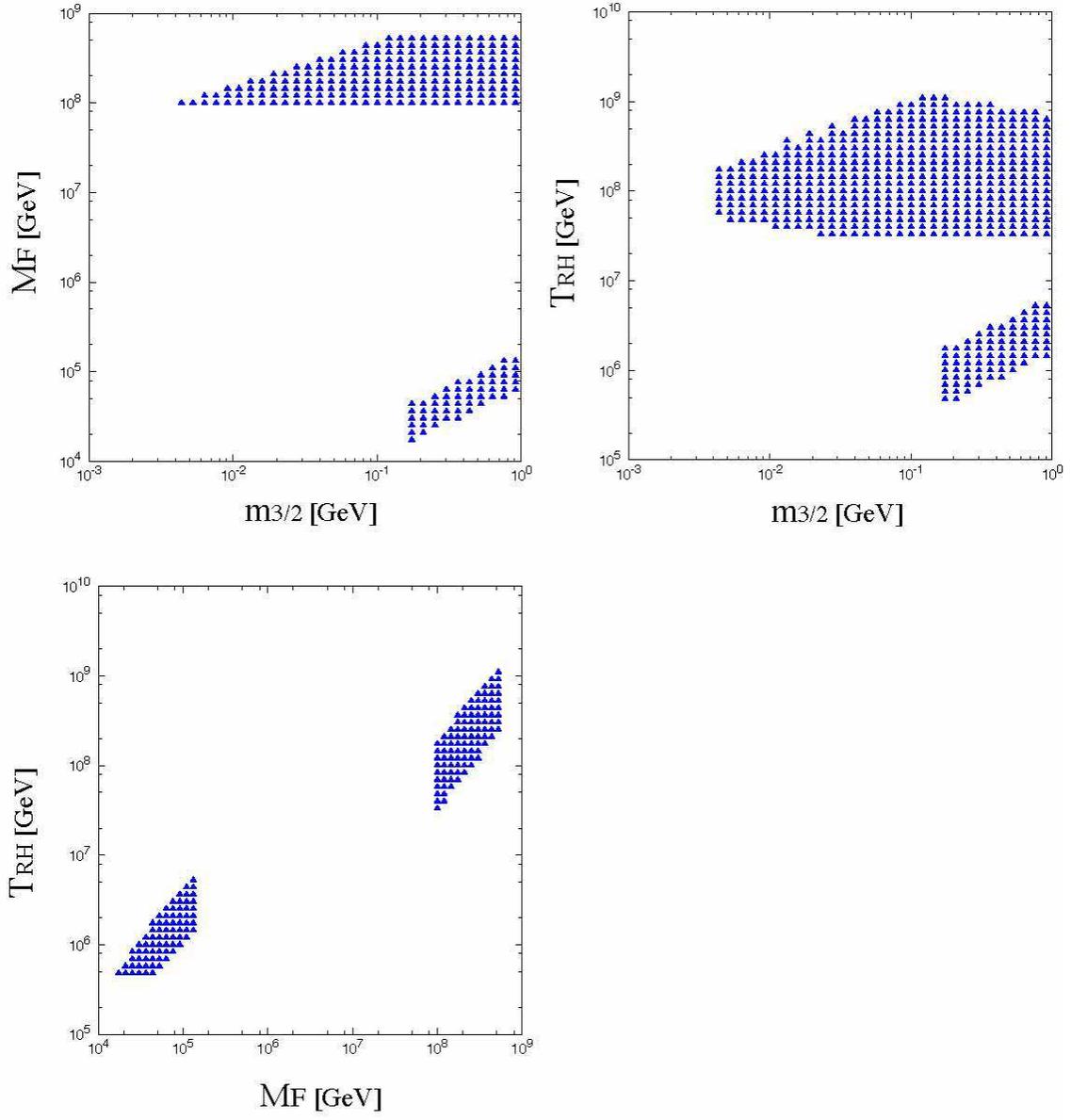}
\caption{\label{fig:dom} 
The allowed regions for $m_{3/2}$, $M_F$ and $T_{RH}$ in the case that
the Q balls dominate the universe are shown. Here we set $f_{(1)}=f_{(2)}=
0.3$, $M=M_G$
and $n=6$.}
\end{figure}

\begin{figure}
\includegraphics[width=15cm]{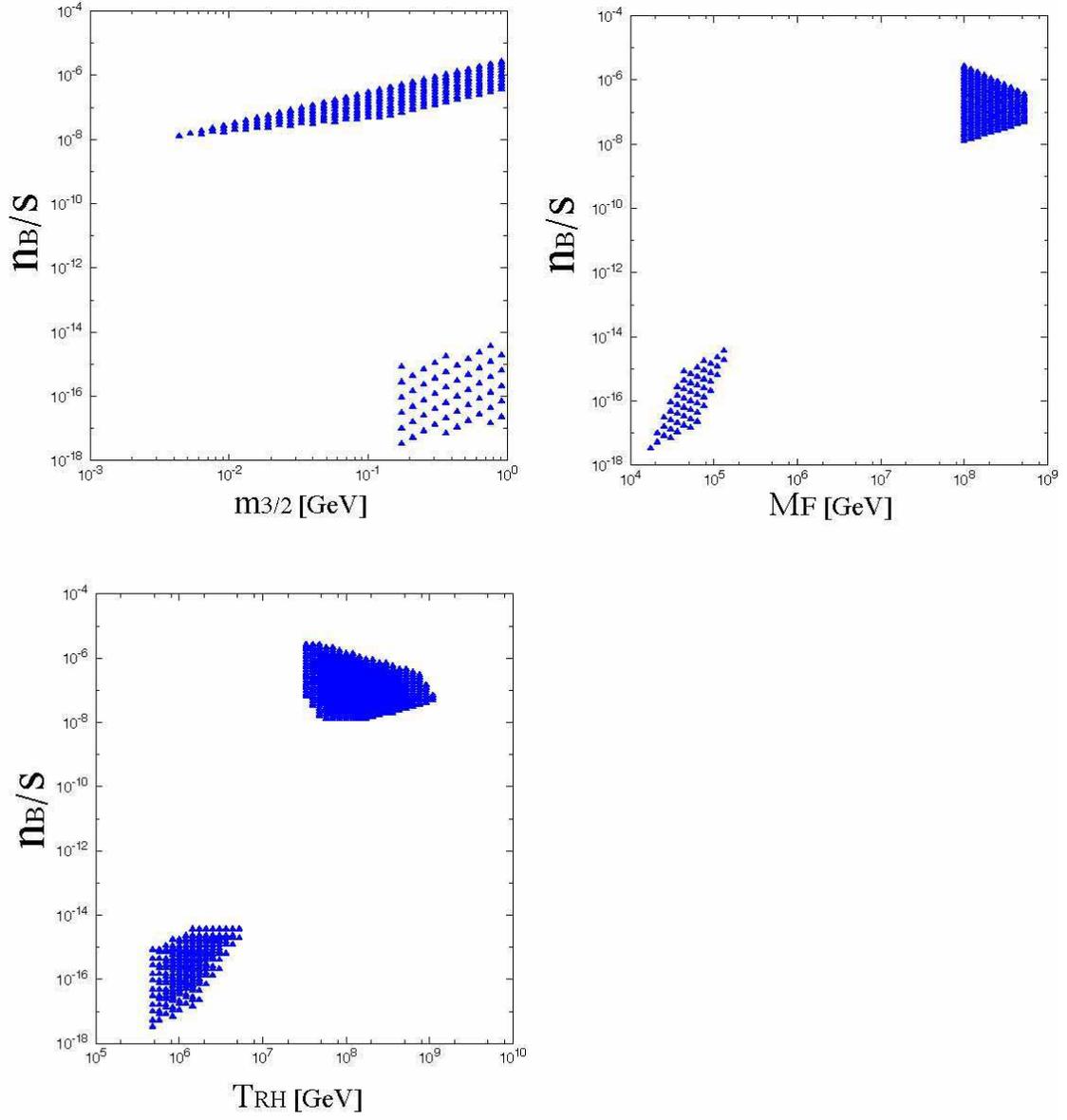} 
\caption{\label{fig:dom2} The
baryon-to-entropy ratio is plotted for the allowed regions in
Fig.~\ref{fig:dom}.}
\end{figure}

To satisfy the requirement that the net baryon number is generated, the
following condition must be satisfied:
\begin{equation}
\label{eq:mtheta}
   M_\theta < H_{osc},
\end{equation}
where 
\begin{eqnarray}
   M_\theta & \simeq & 
   \sqrt{\frac{m_{3/2} \phi_{osc}^{n-2}}{M^{n-3}}},\\
   H_{osc} & \simeq &\left\{
   \begin{array}{ll}
      \ds{\frac{T_{osc}^4}{T_{RH}^2 M_G}} & 
      {\rm for~~} T_{osc} > T_{RH},\\
      &\\
      \ds{\frac{T_{osc}^2}{M_G}} & {\rm for~~} T_{osc} < T_{RH}.
   \end{array}
   \right.
\end{eqnarray}
Then the baryon number is generated as
\begin{equation}
    n_B(t_{osc}) \simeq \frac{1}{M_{\phi}} \frac{m_{3/2}
   \phi_{osc}^n}{M^{n-3}}\,,
\end{equation}
where it should be noted that $M_{\phi} = M_F^2/\phi_{osc}$ is used
instead of the Hubble parameter, because the kick due to the A-term lasts
for $\delta t \sim M_{\phi}^{-1}$, and we also set the numerical factor to
be unity.
The almost all baryon number is absorbed into the subsequently formed Q
balls with the typical charge given as Eq.~(\ref{eq:qcharge})~\footnote{
The value of $\beta$ in Eq.(\ref{eq:qcharge}) might be different in our
case, since the Q-ball formation is delayed compared to the ordinary
scenario. However we make use of this equation, since it is still valid
up to the numerical factor.}.
Hereafter we consider the two cases, depending on whether the Q balls
dominate the universe or not.  If the Q balls dominates the universe, they
must be unstable and decay before the epoch of big bang nucleosynthesis
(BBN). Thus, the mass per unit charge and the decay temperature are
constrained as
\begin{eqnarray}
   M_F Q^{-\frac{1}{4}} &>& 1 {\rm GeV},\\
   T_d  = \left(\frac{\pi^2 g_*}{90}\right)^{-\frac{1}{4}}
   \sqrt{\frac{M_F M_G}{48 \pi Q^{\frac{5}{4}}}} &>& 10 {\rm MeV}.
\end{eqnarray}
Since the universe is reheated again, the gravitino problem is alleviated
in this case. Accordingly the constraints Eqs.(\ref{eq:trap}) and
(\ref{eq:con-u2}) with $\gamma$ given as Eq.~(\ref{eq:dilution}) are
adopted. The baryon-to-entropy ratio is given as
\begin{eqnarray}
   \frac{n_B}{s} &=& \frac{3}{4} \frac{n_B(t_{osc}) T_d}{M_F^4},
   \non\\
   &\sim& \left\{
   \begin{array}{ll}
      \ds{30 \times \left(\frac{g_*}{10}\right)^{-\frac{1}{4}}
      m_{3/2}^{-n+\frac{5}{2}}
      M_F^{2(n-3)} M^{-(n-3)} M_G^{\frac{1}{2}}} 
      & {\rm for~~} \ds{m_{3/2} >
      \left(M_F^{2n-4} M^{-(n-3)}\right)^{\frac{1}{n-1}}},\\ 
      &\\ \ds{30\times\left(\frac{g_*}{10}\right)^{-\frac{1}{4}} 
      m_{3/2} M_F^{-\frac{n}{n-1}}
      M^{-\frac{(n-3)}{2(n-1)}} M_G^{\frac{1}{2}}} 
      & {\rm for~~} \ds{m_{3/2} <
      \left(M_F^{2n-4} M^{-(n-3)}\right)^{\frac{1}{n-1}}},
   \end{array}
   \right.
\end{eqnarray}
The allowed regions for $m_{3/2}$, $M_F$ and $T_{RH}$ satisfying the above
constraints are shown in Fig.\ref{fig:dom} and \ref{fig:dom2}, where we
set $f_{(1)}=f_{(2)}=0.3$, $M=M_G$ and $n=6$. It shows that the
baryon-to-entropy ratio can be large enough to explain the present
asymmetry.  For instance, it takes the value as
\begin{equation}
   \frac{n_B}{s}=6 \times 10^{-7}  
   \left(\frac{g_*}{10}\right)^{-\frac{1}{4}}
   \left(\frac{m_{3/2}}{100{\rm MeV}}\right)
   \left(\frac{M_F}{5 \times 10^8 {\rm GeV}}\right)^{-\frac{6}{5}}
   \left(\frac{M}{M_G}\right)^{-\frac{3}{10}},
\end{equation}
where $m_{3/2}$ and $M_F$ is assumed to satisfy all the constraints stated
above, so their values must be included in the allowed regions shown in
Fig.\ref{fig:dom} and \ref{fig:dom2}.  Though the above value is a little
larger compared to the expected one $\sim 10^{-10}$, it might be
suppressed by the unknown CP phase $\theta_{osc} $which we assume to be of
the order of unity here. Although $M>M_G$ is possible since the definition
of $M$ in Eq.~(\ref{eq:spnr}) includes a possibly small coupling constant,
it does not work to take larger $M$. This is because the requirement that
Q balls decay before the BBN sets the lower bound for $M_F$ which is a
positive power of $M$, so the allowed regions shown in Fig.\ref{fig:dom}
and \ref{fig:dom2} would disappear for $M \gg M_G$.

\begin{figure}
\includegraphics[width=15cm]{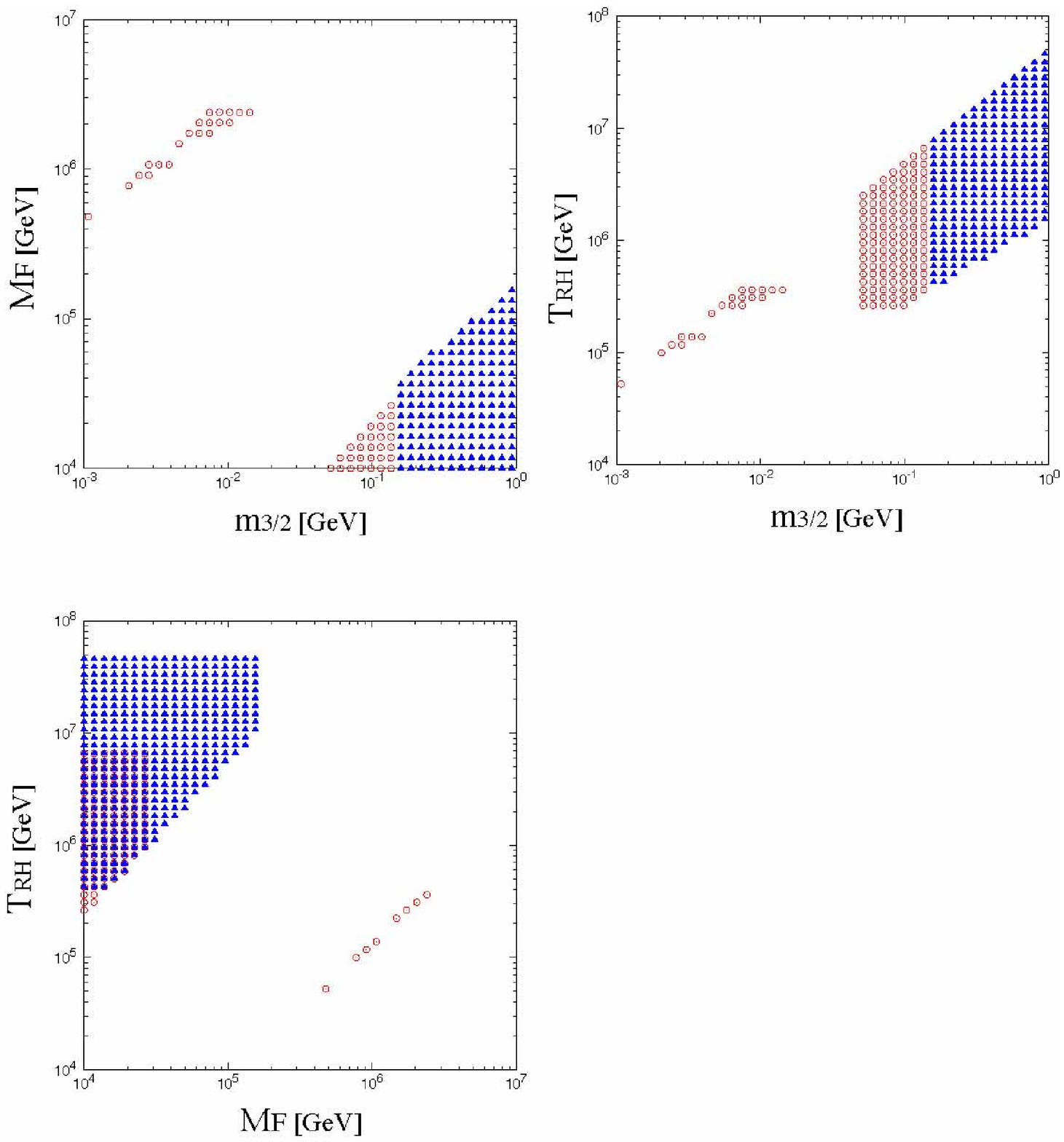} \caption{\label{fig:sub} The
allowed regions for $m_{3/2}$, $M_F$ and $T_{RH}$ in the case that the Q
balls do not dominate the universe are shown.  Here we set
 $f_{(1)}=f_{(2)}=0.3$,
$M=M_G$ and $n=6$. The open cicle and closed triangle denote
stable and unstable Q balls respectively.}
\end{figure}

\begin{figure}
\includegraphics[width=15cm]{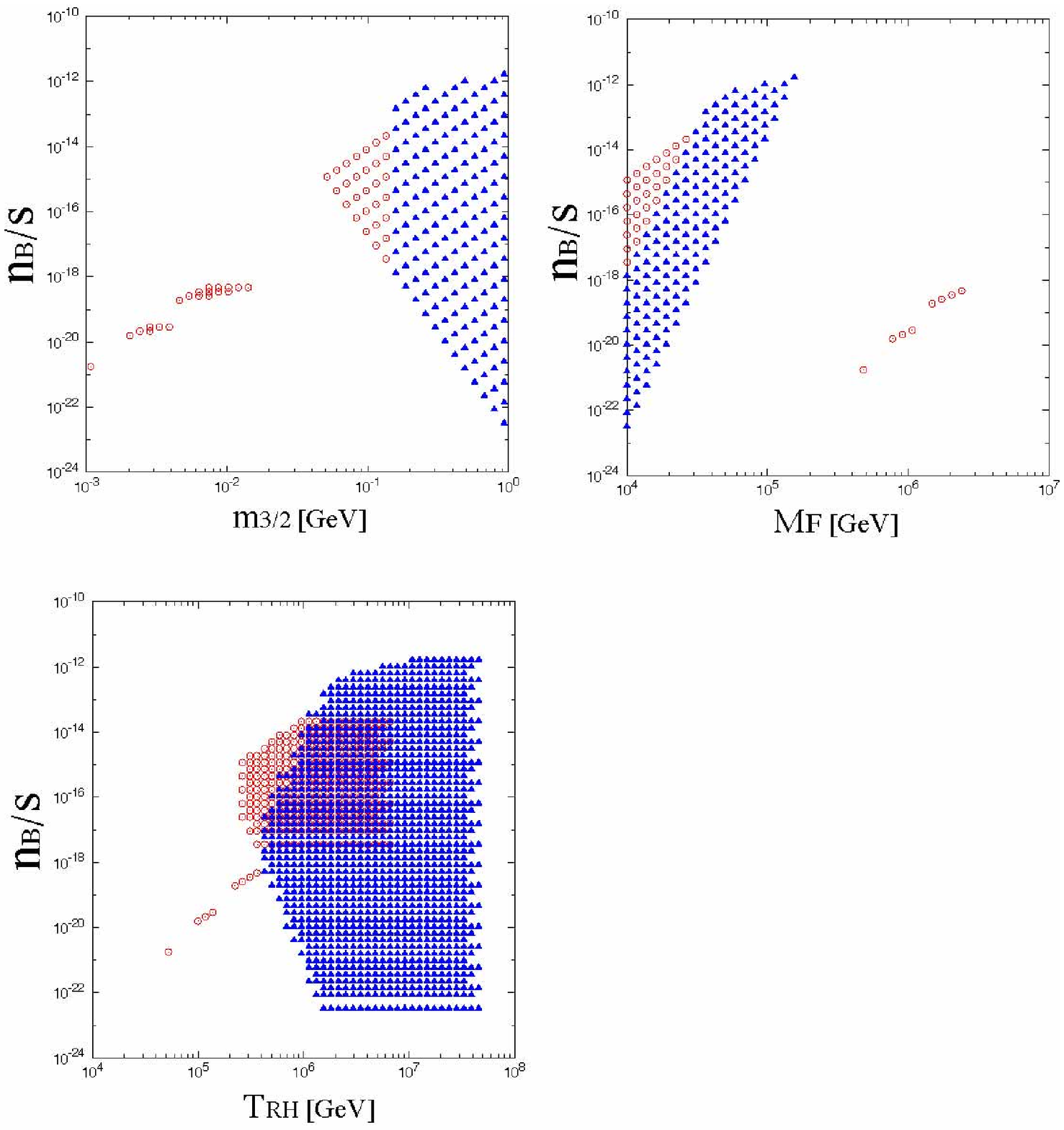} 
\caption{\label{fig:sub2} The baryon-to-entropy
ratio is plotted for the allowed regions in Fig.~\ref{fig:sub}.
The open cicle and closed triangle denote stable and unstable Q balls
respectively. 
}
\end{figure}

\begin{figure}
\includegraphics[width=15cm]{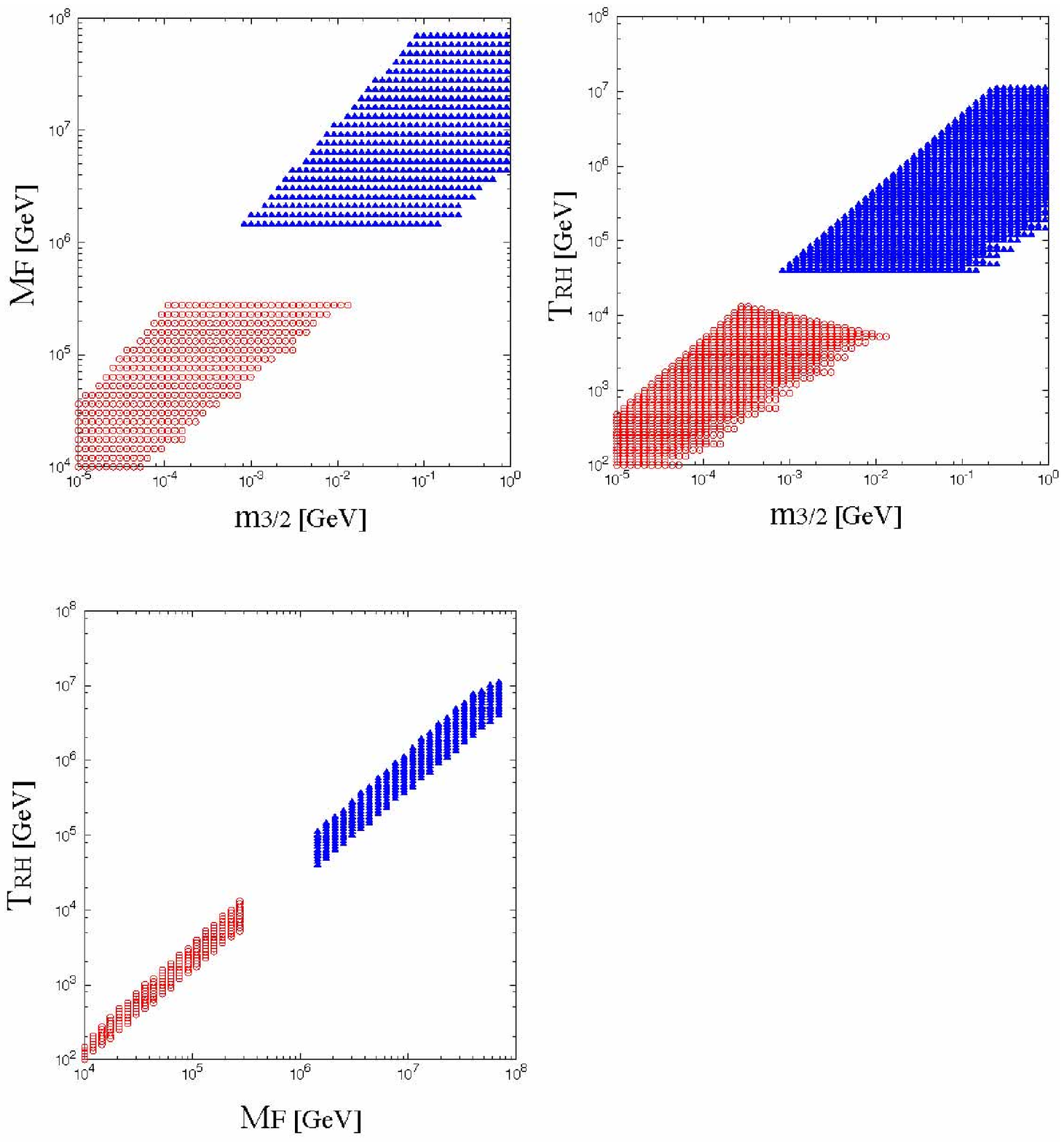} \caption{\label{fig:sub16} The
allowed regions for $m_{3/2}$, $M_F$ and $T_{RH}$ in the case that the Q
balls do not dominate the universe are shown.  Here we set
 $f_{(1)}=f_{(2)}=0.3$,
$M=10^{16}$ GeV and $n=6$. The open cicle and closed triangle denote
stable and unstable Q balls respectively.}
\end{figure}

\begin{figure}
\includegraphics[width=15cm]{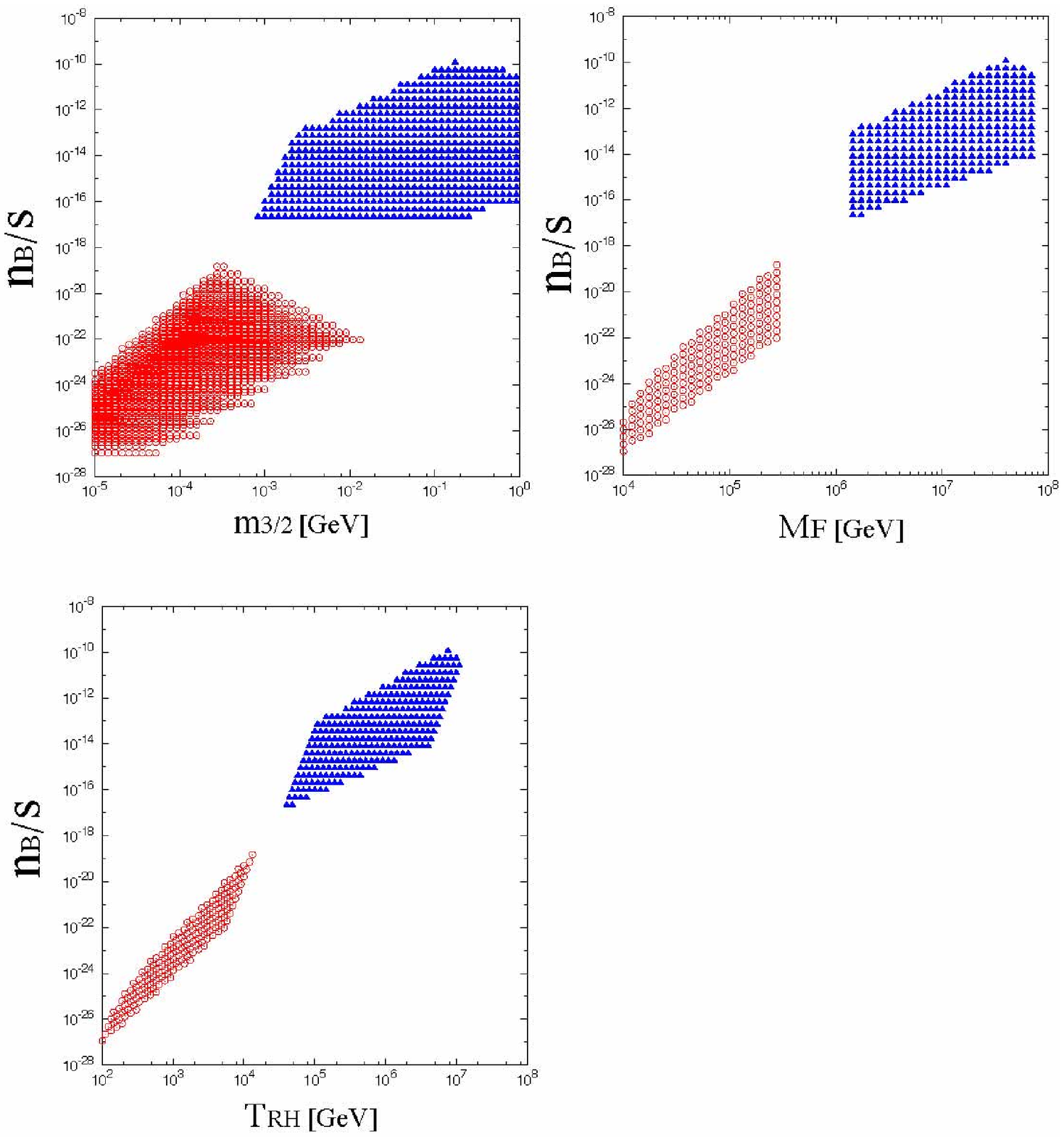} 
\caption{\label{fig:sub16-2} The baryon-to-entropy
ratio is plotted for the allowed regions in Fig.~\ref{fig:sub16}.
The open cicle and closed triangle denote stable and unstable Q balls
respectively. 
}
\end{figure}

Next we consider the case that the energy of Q balls is subdominant
before the BBN. If Q balls are unstable, they must decay before the BBN,
otherwise their decay product might destroy the light elements and
spoil the success of the BBN. If Q balls are stable, the resultant
baryon asymmetry is suppressed by $\Delta Q/Q$, where $\Delta Q$ is the
evaporated charge. Also the abundance of  remnant Q balls are
constrained in order not to overclose the universe. Namely,
\begin{equation}
   \Omega_Q = \frac{M_F Q^{-1/4}}{m_n} \frac{\Omega_B}{\varepsilon}
   \frac{Q}{\Delta Q} < 1,
\end{equation}
where $m_n$ is the nucleon mass, and $\Omega_{Q(B)}$ is the ratio of the
energy density of Q balls (or baryon) to the critical density. 
The baryon-to-entropy ratio is expressed as
\begin{equation}
   \frac{n_B}{s} \simeq
   \left(\frac{45}{2 \pi^2 g_*}\right) \frac{n_B(t_{osc})}
   {T_{osc}^3}\times \kappa_1 \times \kappa_2,
\end{equation}
where 
\begin{eqnarray}
   \kappa_1 &\equiv& \left\{
   \begin{array}{ll}
      1 & {\rm ~~~~~for~~} T_{osc} < T_{RH},\\
      &\\
      T_{RH}^5/T_{osc}^5 & {\rm ~~~~~for~~}T_{osc} > T_{RH},
   \end{array}
   \right. \\
   &&\non\\
   \kappa_2 &\equiv& \left\{
   \begin{array}{ll}
      1 & {\rm ~~~~~for~~unstable~~Q~~balls},\\
      &\\
      \Delta Q/Q & {\rm ~~~~~for~~stable~~Q~~balls}.
   \end{array}
   \right.
\end{eqnarray}
Since the reheating temperature is now constrained by the gravitino
problem, the constraints Eqs.(\ref{eq:trap}) and (\ref{eq:con-u2}) with
$\gamma = 1$ are adopted.  The allowed regions for $m_{3/2}$, $M_F$ and
$T_{RH}$ satisfying the above constraints are shown in Fig.\ref{fig:sub}
and \ref{fig:sub2}, where we set $f_{(1)}=f_{(2)}=0.3$, $M=M_G$ and
$n=6$. It shows that the baryon-to-entropy ratio can be as large as $\sim
10^{-12}$, which is a little smaller than the favored value $\sim
10^{-10}$. However, it can be cured by taking smaller $M$, the GUT scale
for instance.  Fig.\ref{fig:sub16} and \ref{fig:sub16-2} represent the
allowed regions for $m_{3/2}$, $M_F$ and $T_{RH}$ in the case of
$f_{(1)}=f_{(2)}=0.3$, $M=10^{16}$ GeV and $n=6$.  It shows that the
baryon-to-entropy ratio can be now as large as $10^{-10}$ for unstable Q
balls, although it is too small to explain the present asymmetry for
stable Q balls. For instance, it is given as
\begin{equation}
   \frac{n_B}{s} \sim 3 \times 10^{-11}  \left(\frac{g_*}{200}\right)^{-1}
   \left(\frac{m_{3/2}}{100{\rm MeV}}\right)
   \left(\frac{M_F}{2\times 10^7 {\rm GeV}}\right)^{-\frac{36}{5}}
   \left(\frac{T_{RH}}{ 10^6 {\rm GeV}}\right)^{5}
   \left(\frac{M}{10^{16} {\rm GeV}}\right)^{\frac{6}{5}},
\end{equation}
for unstable Q balls.


\section{Conclusion}
\label{sec:con}

%
We have studied whether the AD baryogenesis works in the gauge-mediated
SUSY breaking models, if the AD field is trapped by the negative thermal
log potential which is known to be ubiquitous in the MSSM.  The most
striking feature is that the Hubble parameter becomes much smaller than
the mass scale of the radial component of the AD field, during the
trap. Then, 
the amplitude of the AD field decreases so slowly that the baryon number
is not fixed even after the onset of radial oscillation.  A further
important point is that the AD field experiences spatial instabilities and
deforms into Q balls.  Therefore the baryon asymmetry can be washed out
due to the random kick in the phase direction during the process of
Q-ball formation.  Indeed, it has been shown that the net baryon number
vanishes if the Hubble parameter is smaller than the mass scale of phase
direction, since the randomly generated baryon number surpasses the baryon
number just before the nonlinear stage begins. This is always the case if
the Hubble-induced A-term exists.
Also we confirmed the chaotic behavior of the baryon number with use of
numerical calculation. However, the Hubble-induced A-term does not
necessarily exist.  In fact, we have found that it is possible to generate
the right amount of the baryon asymmetry in the absence of the
Hubble-induced A-term, if the Q balls are unstable.  Thus, it should be
emphasized that the AD baryogenesis is still a viable candidate, even if
both the thermal effect and Q-ball formation are fully taken into account.

\subsection*{ACKNOWLEDGMENTS}
F.T. thanks Masahide Yamaguchi for useful discussion and comments.

\end{document}